\documentclass[final]{svjour2}
\usepackage{graphicx}
\usepackage{rotating}
\usepackage{amssymb}
\usepackage{mathptmx}
\usepackage[numbers]{natbib}

\usepackage{bm}

\def\up{\uparrow}
\def\dwn{\downarrow}

\def\lesssim{\ \raise.3ex\hbox{$<$}\kern-0.8em\lower.7ex\hbox{$\sim$}\ }
\def\gesim{\ \raise.3ex\hbox{$>$}\kern-0.8em\lower.7ex\hbox{$\sim$}\ }

\makeatletter
\journalname{Journal of Low Temperature Physics}

\bibpunct{}{}{,}{s}{}{,}

\begin{document}

\newcommand{\hdblarrow}{H\makebox[0.9ex][l]{$\downdownarrows$}-}
\title{Magnetic properties and strong-coupling corrections in an ultracold Fermi gas with population imbalance}
\author{T. Kashimura \and R. Watanabe \and Y. Ohashi}

\institute{Department of Physics, Keio University\\
Yokohama, 223-8522, Japan\\
Tel.:+81-45-566-1811\\ 
Fax:+81-45-566-1811\\
\email{t.kashimura@a3.keio.jp}
}
\date{7.1.2012}
\maketitle
\begin{abstract}
We investigate magnetic properties of an ultracold Fermi gas with population imbalance. In the presence of population imbalance, the strong-coupling theory developed by Nozi\`eres and Schmitt-Rink (which is frequently referred to as the NSR theory, or Gaussian fluctuation theory) is known to give unphysical results in the BCS-BEC crossover region. We point out that this problem comes from how to treat pseudogap effects originating from pairing fluctuations and many-body corrections to the spin susceptibility. We also clarify how to overcome this problem by including higher order fluctuations beyond the ordinary $T$-matrix theory. Calculated spin susceptibility based on our extended $T$-matrix theory agrees well with the recent experiment on a $^6$Li Fermi gas.

\end{abstract}
\section{Introduction}
High controllability of ultracold Fermi gases enables us to study various many-body phenomena in a systematic manner\cite{Giorgini,Bloch}. Using a tunable pairing interaction associated with a Feshbach resonance, we can study properties of Fermi superfluids from the weak-coupling regime to the strong-coupling limit in a unified manner\cite{Chin}. Introducing population imbalance to a superfluid Fermi gas\cite{Zwierlein, Partridge}, one can simulate Fermi superfluids discussed in various research fields, such as metallic superconductivity under an external magnetic field, and color superconductivity in a dense quark matter\cite{LiuWilczek}. The population imbalance naturally induces mismatch of Fermi surfaces between (pseudo)spin-$\uparrow$ and (pseudo)spin-$\downarrow$ components, which suppresses the conventional $s$-wave superfluid state\cite{Clogston}. In this case, other pairing states, such as the Fulde-Ferrell-Larkin-Ovchinnikov (FFLO) state\cite{FF,LO} and the Sarma state, have been predicted\cite{Sheehy}. 
\par
When we deal with both the tunable interaction and population imbalance at the same time, pairing fluctuations are crucial in determining the phase diagram in terms of the interaction strength, polarization, and temperature. Even in the weak-coupling regime, it has been pointed out\cite{Shimahara,Ohashi} that the FFLO phase is unstable against pairing fluctuations in the absence of a lattice potential. However, despite the importance of strong-coupling effects, polarized Fermi gases have been so far mainly examined within the mean-field level, so that roles of pairing fluctuations in this system have not been fully understood yet. One of the crucial reasons for this  current situation is the breakdown of the strong-coupling theory developed by Nozi\`eres and Schmitt-Rink (NSR)\cite{NSR} in the presence of population imbalance\cite{Liu, Parish}. While the NSR theory has succeeded in explaining the BCS (Bardeen-Cooper-Schrieffer)-BEC (Bose-Einstein condensation) crossover physics in unpolarized Fermi gases, it gives negative spin susceptibility in the presence of population imbalance\cite{Liu}, which is, however, forbidden thermodynamically\cite{Sewell}. 
\par
To resolve the above mentioned problem that the NSR theory possesses, we investigate a polarized Fermi gas in the BCS-BEC crossover region. We point out that one must carefully treat the pseudogap phenomenon associated with pairing fluctuations, as well as many-body corrections to the spin fluctuations in the presence of population imbalance. To overcome this problem, the ordinary $T$-matrix theory (which has been also frequently applied to the unpolarized case) is found to be still not enough. Further extending the $T$-matrix theory to include higher order fluctuation effects, we obtain the expected positive spin susceptibility in the entire BCS-BEC crossover region. The calculated spin susceptibility agrees well with the recent experiment on a $^6$Li Fermi gas. Using this extended $T$-matrix theory, we also determine the superfluid phase transition temperature $T_{\rm c}$ in a polarized Fermi gas. 
\par
\section{Formulation}
\par
We consider a two-component Fermi gas with population imbalance ($N_\uparrow>N_\downarrow$, where $N_\sigma$ is the number of atoms with pseudospin $\sigma=\uparrow, \downarrow$). Since recent experiments are all using a broad Feshbach resonance, we employ the single-channel BCS model, described by the Hamiltonian, 
\begin{eqnarray}
H = \sum_{\bm{p}, \sigma} \xi_{\bm{p}, \sigma} c^\dagger_{\bm{p}, \sigma} c_{\bm{p}, \sigma}
- U  \sum_{\bm{p}, \bm{p}', \bm{q}} c^\dagger_{\bm{p}+\bm{q}/2, \up} c^\dagger_{-\bm{p}+\bm{q}/2, \dwn} c_{\bm{-p}'+\bm{q}/2, \dwn} c_{\bm{p}'+\bm{q}/2, \up}.
\label{H}
\end{eqnarray}
Here, $c^\dagger_{\bm{p},\sigma}$ is the creation operator of a Fermi atom with momentum $\bm{p}$ and pseudospin $\sigma$. $\xi_{\bm{p},\sigma} \equiv \epsilon_{\bm{p}}-\mu_\sigma= p^2/2m-\mu_\sigma$ is the kinetic energy, measured from the chemical potential $\mu_\sigma$ (where $m$ is an atomic mass). When we write the chemical potential as $\mu_\sigma=\mu+\sigma h$, the Hamiltonian in Eq. (\ref{H}) effectively describes a system under an external magnetic field $h$. The pairing interaction $-U$ ($<0$) is assumed to be tunable by a Feshbach resonance, which is related to the $s$-wave scattering length $a_s$ as $4 \pi a_s/m = -U/\left[1-U\sum_{\bm{p}}^{\omega_{\rm c}} (m/p^2) \right]$ (where $\omega_{\rm c}$ is a cutoff energy). In this scale, the weak-coupling BCS regime and the strong-coupling BEC regime are, respectively, characterized by $(k_{\rm F} a_s)^{-1} \lesssim -1$ and $(k_{\rm F} a_s)^{-1} \gesim 1$ (where $k_{\rm F}$ is the Fermi momentum). The region $-1 \lesssim (k_F a_s)^{-1} \lesssim 1$ is called the crossover region.
\par
Strong-coupling effects are conveniently described by the self-energy $\Sigma_{\bm{p}, \sigma}(i\omega_n)$ in the single-particle thermal Green's function, 
\begin{equation}
G_{\bm{p},\sigma} (i\omega_n) = [G^0_{\bm{p},\sigma}(i\omega_n)^{-1}-\Sigma_{\bm{p}, \sigma}(i\omega_n)]^{-1},
\label{GF}
\end{equation} 
where $\omega_n$ is the fermion Matsubara frequency, and $G^0_{\bm{p},\sigma}(i\omega_n)=[i \omega_n-\xi_{\bm{p},\sigma}]^{-1}$ is the non-interacting Green's function. In this paper, we take the self-energy as
\begin{equation}
\Sigma_{\bm{p}, \sigma}(i\omega_n) = T \sum_{\bm{q}, i\nu_n} \Gamma_{\bm q}(i\nu_n) G_{\bm{q-p}, -\sigma} (i\nu_n-i\omega_n).
\label{SE}
\end{equation}
Here, $\nu_n$ is the boson Matsubara frequency. The particle-particle scattering matrix $\Gamma_{\bm q}(i\nu_n) = -U/[1-U \Pi_{\bm{q}}(i\nu_n)]$ involves fluctuation contributions described by the sum of ladder-type diagrams\cite{Tsuchiya}, where 
\begin{equation}
\Pi_{\bm{q}}(i\nu_n) = T \sum_{\bm{p},i \omega_n} G^0_{\bm{p+q}/2,\uparrow} (i\nu_n+i\omega_n) G^0_{\bm{-p+q}/2,\downarrow} (-i\omega_n)
\end{equation}
is the lowest-order pair-propagator.
\par
Th self-energy in the ordinary $T$-matrix theory is given by Eq. (\ref{SE}) where $G_{{\bm p},\sigma}(i\omega_n)$ is replaced by $G^0_{{\bm p}\sigma}(i\omega_n)$, as
\begin{equation}
\Sigma^{\rm TMA}_{\bm{p}, \sigma}(i\omega_n) = T \sum_{\bm{q}, i\nu_n} \Gamma_{\bm q}(i\nu_n) G^0_{\bm{q-p}, -\sigma} (i\nu_n-i\omega_n).
\label{SET}
\end{equation}
The NSR theory also uses Eq. (\ref{SET}), where Eq. (\ref{GF}) is expanded to O($\Sigma^{\rm TMA}_{\bm p}(i\omega_n)$), as
\begin{equation}
G^{\rm NSR}_{{\bm p}, \sigma} (i\omega_n)=G_{{\bm p}, \sigma}^0(i\omega_n)+G_{{\bm p}, \sigma}^0(i\omega_n)\Sigma_{{\bm p}, \sigma}^{\rm TMA}(i\omega_n)G_{{\bm p}, \sigma}^0(i\omega_n).
\label{NSR}
\end{equation}
For clarify, we call the strong-coupling theory using the self-energy in Eq. (\ref{SE}) the {\it extended} $T$-matrix theory in this paper. 
\par
The superfluid phase transition temperature $T_{\rm c}$ is determined from the Thouless criterion
\begin{equation}
\Gamma^{-1}_{\bm q}(i\nu_n=0)=0.
\label{ThoulessC}
\end{equation}
When the highest $T_{\rm c}$ is obtained at ${\bm q}=0$, the ordinary uniform superfluid phase is realized. The FFLO state, on the other hand, corresponds to the case of ${\bm q}\ne 0$. However, since the latter state is known to be unstable against pairing fluctuations in the absence of a background lattice potential\cite{Shimahara,Ohashi}, we only consider the case of ${\bm q}=0$ in this paper. As usual, we solve Eq. (\ref{ThoulessC}), together with the number equations
\begin{equation}
N_{\sigma=\uparrow,\downarrow} = T \sum_{\bm{p},i\omega_n} G_{\bm{p},\sigma} (i\omega_n).
\label{Ns}
\end{equation}
\par
\begin{figure}
\begin{center}
\includegraphics[%
  width=0.65\linewidth,
  keepaspectratio]{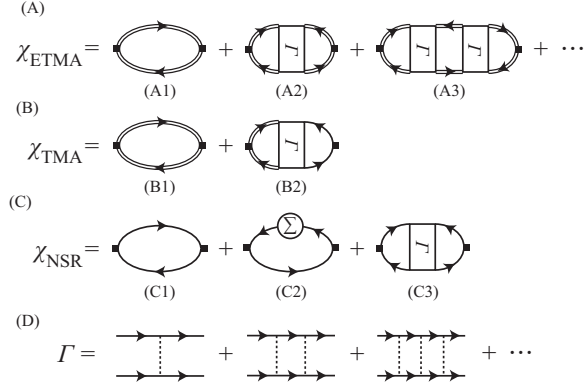}
\end{center}
\caption{Feynman diagrams describing the spin susceptibility $\chi$. (A) $\chi_{\rm ETMA}$: extended $T$-matrix approximation. (B) $\chi_{\rm TMA}$: $T$-matrix approximation. (C) $\chi_{\rm NSR}$: NSR theory. The double lines and single solid lines describe the full Green's function $G$ and the free Green's function $G^0$, respectively.  We note that, the so-called `Aslamazov-Larkin' diagrams vanish identically in the spin susceptibility. (D) scattering matrix $\Gamma$ in the $T$-matrix approximation, where the dashed lines describe the interaction $-U$.
}
\label{fig1}
\end{figure}

\section{Spin susceptibility and many-body corrections}
\par
The spin susceptibility $\chi$ is given by
\begin{equation}
\chi={N_\uparrow-N_\downarrow \over h}\Bigr|_{h\to0}
=T \sum_{\bm{p},i\omega_n,\sigma} \sigma
{\partial G_{\bm{p},\sigma} (i\omega_n) \over\partial h}\Bigr|_{h\to 0}.
\label{SUS}
\end{equation}
Substituting the Green's function given by Eq. (\ref{GF}) into Eq. (\ref{SUS}), we automatically obtain, not only self-energy corrections, but also vertex corrections to the spin susceptibility in a consistent manner. In the  extended $T$-matrix theory (where the self-energy in Eq. (\ref{SE}) is used), many-body corrections to the spin susceptibility ($\equiv\chi_{\rm ETMA}$) are diagrammatically given by the first line in Fig.\ref{fig1}. Here, the first term (A1) involves many-body corrections to $\chi$ through the density of states $\rho(\omega)$. That is, noting that the spin susceptibility $\chi$ is deeply related to $\rho(\omega)$ at the Fermi level ($\omega=0$), when pairing fluctuations induce the pseudogap in $\rho(\omega)$ above $T_{\rm c}$, $\chi$ is also suppressed due to the suppression of $\rho(\omega)$ around $\omega=0$. To understand the role of (A2), (A3) and higher order diagrams in Fig.\ref{fig1}, it is convenient to approximate $\Gamma$ to the lowest-order contribution $-U$. Then, the sum of these diagrams are found to be similar to the random-phase approximation (RPA), as 
\begin{equation}
\chi_{\rm ETMA}\simeq
{{\bar \chi} \over 1+U{\bar \chi}},
\label{SUSETMA}
\end{equation}
where ${\bar \chi}$ is given by the diagram (A1) in Fig.\ref{fig1}. 
\par

\begin{figure}
\begin{center}
\includegraphics[%
  width=0.48\linewidth,
  keepaspectratio]{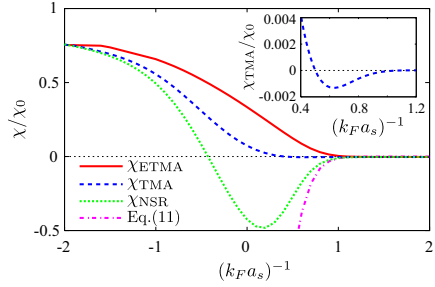}
\end{center}
\caption{(Color online) Spin susceptibility $\chi$ at $T_{\rm c}$, as a function of the interaction strength. $\chi_0$ is the spin susceptibility in a free Fermi gas at $T=0$. The interaction is measured in terms of the inverse scattering length $a_s$. The inset shows $\chi_{\rm TMA}$ magnified around the region where $\chi_{\rm TMA}<0$.}
\label{fig2}
\end{figure}

When one uses the NSR Green's function $G^{\rm NSR}=G^0+G^0\Sigma^{\rm TMA}G^0$ in Eq. (\ref{SUS}), the resulting spin susceptibility ($\equiv\chi_{\rm NSR}$) is diagrammatically described by the third line in Fig.\ref{fig1}. Reflecting that the self-energy is only taken into account to $O(\Sigma^{\rm TMA})$, the diagram (A1) is now approximated to (C1) and (C2). In this regard, we note that the NSR Green's function $G^{\rm NSR}$ is known to overestimate the magnitude of the pseudogap\cite{Tsuchiya}, leading to the unphysical {\it negative} density of states around the Fermi level in the BCS-BEC crossover region. Because of this, the suppression of the spin susceptibility by the pseudogap effect is also overestimated in $\chi_{\rm NSR}$, leading to the negative spin susceptibility, as shown in Fig.\ref{fig2}. In the strong-coupling BEC limit, one finds\cite{Kashimura}
\begin{equation} \label{AFx} 
\chi_{\rm NSR} = \bar{\chi}_0 -\frac{16 \pi a_s}{m} \left( \frac{2mT_{\rm c}^{\rm BEC}}{2\pi} \right)^{\frac{3}{2}}  \zeta \left( \frac{3}{2} \right) \frac{\partial^2 N^0_\up }{\partial h^2 }\Bigg|_{h=0}, 
\end{equation}
where $\bar{\chi}_0 = (2T)^{-1} \sum_{\bm{p}} \cosh^{-2} ({\xi_{\bm{p}}/2T })$
 is the susceptibility of a free Fermi gas. $N_\uparrow^0 = \sum_{\bm{p}} [e^{\beta (\epsilon_{\bm{p}}-\mu-h)}+1]^{-1} $ and $T_{\rm c}^{\rm BEC}=0.218T_{\rm F}$ is the superfluid phase transition temperature in the BEC limit (where $T_{\rm F}$ is the Fermi temperature). The second term in Eq. (\ref{AFx}) just arises from the diagram (C2) in Fig. \ref{fig1}, which is related to fluctuation corrections to the density of states\cite{Varlamov}. As shown in Fig.\ref{fig2}, this term is the origin of the negative spin susceptibility in the BEC regime.
\par
The problem of the negative density of states is absent in the ordinary $T$-matrix theory (where Eq. (\ref{SET}) is used)\cite{Tsuchiya}. In this case, the (A1)-type diagram in Fig.\ref{fig1} is correctly included as (B1) in the second line of Fig.\ref{fig2}. However, when we evaluate $\chi$ based on this theory ($\equiv\chi_{\rm TMA}$), although the situation becomes much better than the NSR result, $\chi_{\rm TMA}$ still becomes negative for $0.5 \lesssim (k_F a_s)^{-1} \lesssim 1.1$. (See the inset in Fig.\ref{fig2}). This is because the RPA-type series in the extended $T$-matrix theory is truncated at the first order ((B2) in Fig.\ref{fig2}) in the ordinary $T$-matrix theory. Evaluating $\chi_{\rm TMA}$ by approximating $\Gamma$ to $-U$, one obtains
\begin{equation}
\chi_{\rm TMA}\simeq{\bar \chi}-\bar{\chi}_0 U{\bar \chi}.
\label{SUSTMA}
\end{equation}
While Eq.(\ref{SUSETMA}) is always positive, Eq.(\ref{SUSTMA}) becomes negative when $U{\bar \chi_0}>1$. 
\par
The above analyses clearly indicate that one needs to carefully treat the pseudogap effect and the RPA-type many-body corrections to the spin susceptibility in the presence of population imbalance. Since they are both correctly treated in our extended $T$-matrix theory, the calculated spin susceptibility $\chi_{\rm ETMA}$ is positive in the entire BCS-BEC crossover. As shown in Fig.\ref{fig2}, $\chi_{\rm ETMA}$ gradually decreases with increasing the interaction strength, and almost vanishes when $(k_{\rm F}a_s)^{-1}\gesim 1$, because the spin degrees of freedom almost dies out due to the formation of tightly binding singlet pairs in the BEC regime. 
\par
\begin{figure}
\begin{center}
\includegraphics[%
  width=0.4\linewidth,
  keepaspectratio]{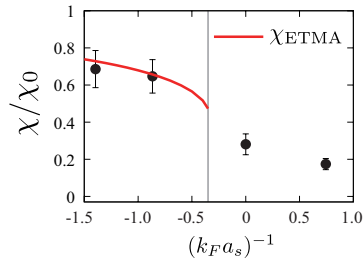}
\end{center}
\caption{(Color online) Comparison of calculated spin susceptibility $\chi_{\rm ETMA}$ (solid line) with the recent experiment on a $^6$Li Fermi gas (where the spin susceptibility is measured by using {\it in situ} imaging of dispersive speckle patterns) (solid circles)\cite{Sanner}. In this experiment, the temperature is fixed at the value of $T_{\rm c}$ at $(k_Fa_s)^{-1}=-0.35$. Thus, the right side of the vertical line is in the superfluid state.}
\label{fig3}
\end{figure}

We compare our theoretical result with the recent experiment on $^6$Li\cite{Sanner} in Fig.\ref{fig3}. The calculated spin susceptibility $\chi_{\rm ETMA}$ well agrees with the observed susceptibility in the normal state (the left side of the vertical dotted line). We emphasize that there is no fitting parameter in obtaining Fig.\ref{fig3}.
\par
\par
\begin{figure}
\begin{center}
\includegraphics[%
  width=0.4\linewidth,
  keepaspectratio]{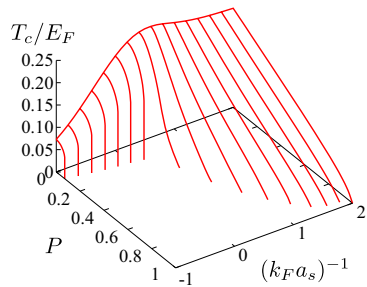}
\end{center}
\caption{(Color online) Superfluid phase transition temperature $T_{\rm c}$ of a Fermi gas with population imbalance.}
\label{fig4}
\end{figure}

So far, we have discussed spin susceptibility, which is obtained in the zero polarization limit ($h\to 0$). Here, we briefly consider the case of finite polarization. Figure \ref{fig4} shows the phase diagram of a polarized Fermi gas in terms of the interaction strength, polarization $P=(N_\up-N_\dwn)/(N_\up+N_\dwn)$, and the temperature. In the weak-coupling BCS regime, $T_{\rm c}$ vanishes at a critical polarization $P_{\rm c}$. The magnitude of $P_{\rm c}$ increases, as one passes through the BCS-BEC crossover region. In the BEC regime, one finds $T_{\rm c} = T_{\rm c}^{\rm BEC} (1-P)^{2/3}$.
\par
We briefly note that, although we have only considered the second order phase transition by using the Thouless criterion in Fig.\ref{fig4}, the first order transition may exist below $P_{\rm c}$\cite{Sheehy,Parish}. To examine the first order phase transition of a polarized Fermi gas, one needs to take into account the possibility of phase separation between the paired molecules and fully polarized Fermi atoms, which remains as our future problem.  
\par
\par
\section{Summary}
To summarize, we have investigated an ultracold Fermi gas with population imbalance. We clarified that the breakdown of the NSR and $T$-matrix theories in the presence of population imbalance can be eliminated by carefully treating the pseudogap effects and the RPA-type many-body corrections to the spin susceptibility. The resulting extended $T$-matrix theory gives the expected positive spin susceptibility in the entire BCS-BEC crossover region. The calculated spin susceptibility well agrees with the recent experiment on a $^6$Li Fermi gas. Since polarized Fermi gases are deeply related to metallic superconductivity under an external magnetic field, our results would be useful for the further development of the physics of polarized Fermi superfluids.

\begin{acknowledgements}
We would like to thank S. Watabe, Y. Endo, D. Inotani, and R. Hanai for useful discussions. Y. O. was supported by Grant-in-Aid for Scientific research from MEXT in Japan (22540412, 23104723, 23500056).
\end{acknowledgements}



\begin{thebibliography}{99}
\bibitem{Giorgini} S. Giorgini, L. Pitaevskii, and S. Stringari, {\it Rev. Mod. Phys.} \textbf{80}, 1215 (2008).
\bibitem{Bloch} I. Bloch, J. Dalibard, and W. Zwerger, {\it Rev. Mod. Phys.} \textbf{80}, 885 (2008).
\bibitem{Chin} C. Chin, R. Grimm, P. Julienne, and E. Tiesinga, {\it Rev. Mod. Phys.} \textbf{82}, 1225 (2010).
\bibitem{Zwierlein} M. W. Zwierlein, A. Schirotzek, C. H. Schunck, and W. Ketterle, {\it Science} \textbf{311}, 492 (2006).
\bibitem{Partridge} G. B. Partridge, W. Li, R. I. Kamar, Y.-A. Liao, and R. G. Hulet, {\it Science} \textbf{311}, 503 (2006). 
\bibitem{LiuWilczek} W. V. Liu and F. Wilczek, {\it Phys. Rev. Lett.} \textbf{90}, 047002 (2003).
\bibitem{Clogston} A. M. Clogston, {\it Phys. Rev. Lett.} \textbf{9}, 266 (1962).
\bibitem{FF} P. Fulde and R. A. Ferrell, {\it Phys. Rev.} \textbf{135}, A550 (1964).
\bibitem{LO} A. I. Larkin and Y. N. Ovchinnikov, {\it Sov. Phys. JETP} \textbf{20}, 762 (1965). 
\bibitem{Sheehy} For review, see D. E. Sheehy and L. Radzihovsky, {\it Ann. Phys. (N.Y.)} \textbf{322}, 1790 (2007).
\bibitem{Shimahara} H. Shimahara, {\it J. Phys. Soc. Jpn.} \textbf{67} 1872 (1998).
\bibitem{Ohashi} Y. Ohashi, {\it J. Phys. Soc. Jpn.} \textbf{71}, 2625 (2002).
\bibitem{NSR} P. Nozi\`eres and S. Schmitt-Rink, {\it J. Low Temp. Phys.} \textbf{59}, 195 (1985). 
\bibitem{Liu} X.-J. Liu and H. Hu, {\it Europhys. Lett.}, \textbf{75}, 364 (2006). 
\bibitem{Parish} M. M. Parish, F. M. Marchetti, A. Lamacraft, B. D. Simons, {\it Nat. Phys.} \textbf{3}, 124 (2007).
\bibitem{Sewell} G. L. Sewell, {\it Quantum Mechanics and its Emergent Macrophysics}, Princeton University Press, 2002. 
\bibitem{Tsuchiya} S. Tsuchiya, R. Watanabe, and Y. Ohashi, {\it Phys. Rev. A}, \textbf{80}, 033613 (2009).
\bibitem{Kashimura} T. Kashimura, R. Watanabe, and Y. Ohashi, in preparation. 
\bibitem{Varlamov} A. A. Varlamov, G. Balestrino, E. Milani, and D. V. Livanov, {\it Adv. Phys.} \textbf{48}, 655 (1999).
\bibitem{Sanner} C. Sanner, E. J. Su, A. Keshet, W. Huang, J. Gillen, R. Gommers, and W. Ketterle, {\it Phys. Rev. Lett.} \textbf{106}, 010402 (2011) .
\end{thebibliography}
\end{document}